\documentstyle[11pt]{article}
\newcommand {\ignore}[1]{}

\newcommand{\Frac}[2]{\frac{\displaystyle #1}{\displaystyle #2}}

\newcommand{\bc}{\begin{center}}
\newcommand{\ec}{\end{center}}

\def\ifmath#1{\relax\ifmmode #1\else $#1$\fi}
\def\3quarter{{\textstyle{3 \over 4}}}

\def\ra{\rightarrow}

\def\lf{\leaders\hbox to 1em{\hss.\hss}\hfill}

\def\21{$SU(2) \ot U(1)$}

\def\sm{\hbox{standard model }}

\def\neu{\hbox{neutrino }}

\def\neus{\hbox{neutrinos }}

\def\eq#1{{eq. (\ref{#1})}}

\def\VEV#1{\left\langle #1\right\rangle}

\def\lsim{\raise0.3ex\hbox{$\;<$\kern-0.75em\raise-1.1ex\hbox{$\sim\;$}}}
\def\gsim{\raise0.3ex\hbox{$\;>$\kern-0.75em\raise-1.1ex\hbox{$\sim\;$}}}

\def\bel{\begin{letter}}
\def\eel{\end{letter}}
\def\beq{\begin{equation}}
\def\eeq{\end{equation}}
\def\bef{\begin{figure}}
\def\eef{\end{figure}}
\def\bet{\begin{table}}
\def\eet{\end{table}}
\def\bea{\begin{eqnarray}}
\def\ba{\begin{array}}
\def\ea{\end{array}}
\def\bi{\begin{itemize}}
\def\ei{\end{itemize}}
\def\ben{\begin{enumerate}}
\def\een{\end{enumerate}}
\def\ra{\rightarrow}
\def\ot{\otimes}

\def\eea{\end{eqnarray}}
\def\apj#1#2#3{          {\it Astrophys. J. }{\bf #1} (19#2) #3}

\def\np#1#2#3{           {\it Nucl. Phys. }{\bf #1} (19#2) #3}
\def\pl#1#2#3{           {\it Phys. Lett. }{\bf #1} (19#2) #3}
\def\pr#1#2#3{           {\it Phys. Rev. }{\bf #1} (19#2) #3}

\def\prl#1#2#3{          {\it Phys. Rev. Lett. }{\bf #1} (19#2) #3}

\def\n.c.#1#2#3{         {\it Nuovo Cim. }{\bf #1} (19#2) #3}
\def\r.n.c.#1#2#3{       {\it Riv. del Nuovo Cim. }{\bf #1} (19#2) #3}

\relax
\def\ot{\otimes}
\def\21{$SU(2) \ot U(1)$}
\def\eq#1{{eq. (\ref{#1})}}
\def\lsim{\raise0.3ex\hbox{$\;<$\kern-0.75em\raise-1.1ex\hbox{$\sim\;$}}}
\def\gsim{\raise0.3ex\hbox{$\;>$\kern-0.75em\raise-1.1ex\hbox{$\sim\;$}}}
\def\sm{\hbox{standard model }}
\def\neu{\hbox{neutrino }}

\def\ra{\rightarrow}
\def\beq{\begin{equation}}
\def\eeq{\end{equation}}
\def\ben{\begin{enumerate}}
\def\een{\end{enumerate}}

\def\VEV#1{\left\langle #1\right\rangle}

\def\np#1#2#3{           {\it Nucl. Phys. }{\bf #1} (19#2) #3}
\def\pl#1#2#3{           {\it Phys. Lett. }{\bf #1} (19#2) #3}
\def\pr#1#2#3{           {\it Phys. Rev. }{\bf #1} (19#2) #3}

\def\prl#1#2#3{          {\it Phys. Rev. Lett. }{\bf #1} (19#2) #3}

\begin{document}
\begin{titlepage}
\pagestyle{empty}
\rightline{FTUV/93-35}
\noindent
\today
\hfill Submitted to Phys. Lett. B\\
\begin{center}
{\bf \Large The KeV Majoron as a Dark Matter Particle}\\
\vskip 0.4cm
{\bf V. Berezinsky }
\footnote{E-mail 40618::BEREZINSKY}\\
INFN, Laboratori Nazionali del Gran Sasso\\
I-67010, Assergi (AQ), Italy\\
and Institute for Nuclear Research, Moscow, Russia\\
{\bf and }\\
{\bf J. W. F. Valle}
\footnote{E-mail VALLE@vm.ci.uv.es - 16444::VALLE}\\
Instituto de Fisica Corpuscular - IFIC/CSIC\\
Dept. de F\'isica Te\`orica, Universitat de Val\`encia\\
46100 Burjassot, Val\`encia, SPAIN\\
\vskip 0.4cm
{\bf ABSTRACT}\\
\end{center}
\vskip 0.1cm
\noindent
We consider a very weakly interacting KeV majoron as a dark matter
particle (DMP), which provides both the critical density $\rho_{cr}
= 1.88 \times 10^{-29} h^{2}$ $g/cm^{3}$ and the galactic scale $M_{gal}$
$\sim m^{3}_{Pl}/m^{2}_{J} \sim 10^{12} M_{\odot} (m_{J}/1 KeV)^{-2}$
for galaxy formation. The majoron couples to the leptons only through
some new "directly interacting particles", called DIPS, and this
provides the required smallness of the coupling constants.
If the masses of these DIPS are greater than the scale $V_s$
characterizing the spontaneous violation of the global lepton
symmetry they are absent at the corresponding phase transition
($T \sim V_s$) and the majorons are produced during the phase
transition, never being in thermal equilibrium during the
history of the universe. In the alternative case
$m_{DIP} < V_{s}$ the majorons can
be for a short period in thermal equilibrium.
This scenario is not forbidden by
nucleosynthesis and gives a reasonable
growth factor for the density fluctuations compatible
with the recent restrictions from the COBE experiment.
It provides as a possible signature the
existence of an X-ray line at $E_\gamma = \frac{m_J}{2}$,
produced by the decay $J \rightarrow \gamma + \gamma$.
A particle physics model which provides the required
smallness of the majoron couplings is described. It
realizes the possibility of the KeV majoron as a DMP
in a consistent way and may also lead to observable
rates for flavour violating decays
such as $\mu \ra e \gamma$ and $\mu \ra 3e$,
testable in the laboratory.

\vfill
\noindent
\end{titlepage}
\setcounter{page}{1}
\pagestyle{plain}

\section{Introduction}

The basic scale among the Universe structures, the galactic mass,
$M_{gal} \sim 10^{12} M_\odot$, needs a weakly interacting particle
with mass $m_{X} \sim (m_{Pl}^{3} / M_{gal})^{1/2} \sim 1$ KeV, where
$m_{Pl} = 1.2 \times 10^{19}$ GeV is the Planck mass. This particle
must be electrically neutral and long-lived,
$\tau_{x} > t_0$ where $t_0 = 2.06 \times 10^{17} h$
is the age of Universe. The majoron can be such a particle.

The KeV majoron has been recently discussed in the paper \cite{1}. The novel
element of this work is that the B-L symmetry is broken
explicitly by the gravitational interaction considered
as dimension 5 operators inversely proportional to the
Planck mass $m_{Pl}$. This makes the majoron massive.
The coupling constants in this model are such that
the majoron is an unstable particle with the lifetime $\tau_{J} \ll t_{0}$.
The basic parameter of the model, the vacuum expectation value $V_{BL}$
of the complex scalar $\sigma$ which splits into a heavy scalar field
$\rho$ and the pseudoscalar majoron $J$. Limits on $V_{BL}$ were
derived from the decay of the majoron, nucleosynthesis and
galaxy formation (the COBE observations).

The cosmological aspects of the problem were studied in a
more detailed way in ref. \cite{2} and especially in ref. \cite{3}.
In particular the novel feature, the majoron
strings, is discussed in the last work.

Some attention has also been given in these papers to the
phenomenology of the long-lived majoron, though no
elementary particle physics model for this case
was specified.

In this letter following \cite{1,2,3} we consider the singlet complex
scalar $\sigma$ with $L \neq 0$ or $B-L \neq 0$ as the "parent" field for
the majoron. The L or B-L global symmetry is broken by $\VEV{\sigma} =
V_{s} / \sqrt{2}$. Further on the shall discuss the case $V_{s} < V_{EW}$
(where $V_{EW} \approx 250$ GeV) to avoid the problem of washing away the
baryon asymmetry \cite{4}, see also the detailed discussion in \cite{2}. As a
result $\sigma$ splits into the real scalar field $\rho$ and the majoron
according to
$\sigma = \frac{1}{\sqrt{2}} (V_{s} + \rho) \exp ( i J / V_{s})$.

This phase transition in the Universe occurs at temperature
$T_{c} \sim V_{s}$, while at
$T < T_{c}$ the majorons did not exist (In the papers \cite{1}
and \cite{2} there is an inaccurate statement that the majoron
acquires mass at $T >> V_{s}$). We assume that the majorons
interact directly with some heavy particles, X, with a "usual" coupling
constant $g_{JXX} \sim 10^{-2} - 10^{-3}$, while the interaction
with the ordinary particles (left-handed neutrinos, $\nu$, and
charged leptos, $\l$)
can only be mediated by X-particles and has effective coupling
constants $h_{\nu}$ and $h_{l}$, which are extremely small.
They may arise either at the tree level or via radiative
corrections as considered in the present paper.
Further on we shall refer to the X-particles as
Directly Interacting Particles (DIPS).

\section{Cosmological Scenario}

There are several possible mechanisms for the production of the majorons
in the early Universe.

The most natural mechanism is the thermal one \cite{1}.
For the case of a long-lived majoron, $\tau_{J} > t_0$ where
$t_0 = 2.06 \times 10^{17} h^{-1}$ sec
is the age of the Universe, the majoron must have
extremely small effective coupling constants with the neutrinos and
leptons. Therefore the majoron can be in thermal equilibrium only
due to the interactions with DIP, such as right neutrinos \cite{1}
or charged particles as in the model considered below. This is
possible for DIP masses smaller than the scale of the phase transition
$m_{DIP} < V_{s}$. Then, at the time just after the phase transition,
$t \sim t_{J}$, the space density of the majorons is
\begin{equation}
n_{J} (t_{J}) / n_{\gamma} (t_{J}) = 1/2
\end{equation}

The second plausible mechanism is connected with majoron strings,
unstable due to evaporation of the majorons \cite{2}.

We would like to add to this list a third mechanism which operates
effectively when $m_{DIP} > V_{s}$. In this case at the moment of
the phase transition $T_{J} \sim V_{s}$, the DIPS are not present,
while the interactions with other particles are too weak to keep
the majoron in thermal equilibrium. The majorons are born in a
phase transition at $t=t_{J}$, each majoron being accompanied
by one $\rho$-boson.

Before the phase transition the electrically neutral part of
the scalar potential connected with $\sigma$- field was
\begin{equation}
\label{sigma}
V(\sigma) = \lambda_{s} (\sigma^{*} \sigma - V_{s}^{2}/2)^{2}
\end{equation}
with $\lambda_{s} \sim 1$.

After phase transition,
\begin{equation}
\sigma = \frac{1}{\sqrt{2}} (\VEV{\sigma} + \rho + i J)
\end{equation}
The potential energy of the $\sigma$ field $V(0)$ is
converted mostly into the energy of the $\rho$ bosons
$\rho_\rho(t_J) \sim \lambda V_s^4/4$ and therefore
the space density of both $\rho$ bosons and majorons
becomes
$n_J (t_J) \sim n_\rho (t_J) \sim \lambda V_s^4/4 m_\rho$.
If we denote $m_\rho = V_s/\xi$, where $\xi \gsim 1$, then using
$n_\gamma \sim T_s^3 \sim V_s^3$
one has for $\lambda \sim 1$ that, at the moment of the
phase transition,
\beq
\frac{ n_J (t_J)}{n_\gamma (t_J)} \sim \xi
\eeq
The density of majorons at $t=t_0$ is given by
\beq
\frac{ n_J (t_0)}{n_\gamma (t_0)}
=
\frac{ n_J (t_J)}{n_\gamma (t_J)}
\frac{43/11}{N_{i}}
\eeq
where $N_i$ is the number of degrees of freedom
at $t=t_J$ ($N_i=427/4$ for the particle content
of the \sm and 43/11 is the ratio of degrees of
freedom $N(\gamma + e + \nu) N (\gamma)/N(\gamma + e)$.

Once produced at the phase transition the
$\rho$-bosons will immediately decay to majorons.
For example, the the width for the $\rho$ decay
to two majorons can be parametrized by
\begin{equation}
\label{rhoJJ}
\Gamma(\rho \rightarrow JJ)=\frac{\sqrt 2 G_F}{32 \pi} M_{\rho}^3 g^2_{\rho JJ}
\end{equation}
where the corresponding coupling $g_{\rho JJ}$ is given in terms
of the relevant electroweak and global symmetry
breaking vacuum expectation values.

Therefore the present energy density of these majorons is given by
\begin{equation}
\frac{ \rho_J (t_0)}{\rho_{cr} (t_0)} =
1.5 \xi h^{-2} m_J
\end{equation}
where $\xi=1/2$ for the thermal mechanism of production
and $\xi >1$ for the production at the phase transition. Here
and everywhere else we $m_J$ is the mass of the majoron in KeV.

The majorons are also coupled (through DIPS) both to \neus
as well as to the charged leptons with effective
coupling constants which we denote by
$h_\nu$ and $h_{l}$, respectively. The coupling
$h_\nu$ determines the majoron lifetime as
\begin{equation}
\tau_{J \ra \nu \nu } = \frac{16 \pi}{ h_{\nu}^{2} m_{J}}
\end{equation}
In order to be DMP the majoron lifetime must be longer
than the age of the universe. This translates into
a constraint on its effective coupling $h_\nu$
\begin{equation}
\label{21}
h_{\nu} \leq 1.3 \times 10^{-17}
(\frac{ m_{J} }{1 KeV})^{-1/2} h
\end{equation}

On the other hand the couplings to charged leptons, $h_{l}$,
are constrained by $J \ra \gamma \gamma$ decays.
Indeed, in the most general case the current connected
with the global symmetry whose spontaneous violation
gives rise to the majoron is anomalous, and therefore
the majoron is coupled to photons through the
electromagnetic anomaly of this symmetry.
The effective Lagrangean has a form
\begin{equation}
{\cal L}_{int}^l = g_{J \gamma \gamma}^l
J F_{\mu\nu} \widetilde{F_{\mu\nu}},
\end{equation}
where $g_{J \gamma \gamma}^l$ arises from the diagrams shown
in Fig. 1. The effective triangle loop gives
\begin{equation}
\label{17}
g_{J \gamma \gamma}^l =
\frac{\alpha_{em}}{2\pi}
\frac{h_l}{m_l}
\end{equation}
and the majoron lifetime
\begin{equation}
\tau_{J \gamma \gamma} = \frac{64 \pi^{3}}{\alpha_{em}^{2} h^{2}_{l}}
\left( \frac{m_l}{m_{J}} \right)^2
\frac{1}{m_{J}}
\end{equation}
The contributions from two-loop diagrams
mediated by electrically charged scalar bosons,
as in Fig. 1(b), may be of similar magnitude.

The value of $h_{l}$ can be roughly constrained by
the condition that the energy density of the produced
photons must not exceed the observed energy density
($\sim 10^{-5} \rm{eV/cm^3}$) for $E \lsim 1$ KeV.
Using
$\omega_{X} \sim \Frac{ \rho_{cr} t_0 } { \tau_{J \gamma \gamma} }$
one obtains
\begin{equation}
\label{22}
h_\tau \lsim 6 \times 10^{-13} h^{-1} (\frac{ m_{J}}{1 KeV})^{-3/2}
\end{equation}
Notice that for the case $h_l \sim m_l$ (as in the model discussed
below) the coupling to electrons
\begin{equation}
h_e = 1.7 \times 10^{-16}
\end{equation}
is much smaller than the observational limit from
stellar cooling $h_e \lsim 3 \times 10^{-13}$ \cite{Dear}.

We now turn to other cosmological requirements
for the majoron to be DMP, beyond those given
in \eq{21} and \eq{22}.

The restriction imposed by big-bang nucleosynthesis
has already been discussed in refs \cite{2,3}. The number of
degrees of freedom associated to the majorons at the
time of the nucleosynthesis is smaller than the
observational limit
\begin{equation}
N = \frac{43/11}{N_{i}} = 3.7 \times 10^{-2}
\end{equation}

Let us now discuss the problem of galaxy formation in the
light of the COBE results. The basic scale for the majoron
corresponds to the galactic Jeans mass
\begin{equation}
M_{gal} \approx m^{3}_{Pl}/m^{2}_{J}
\sim 1.6 \times 10^{12} M_{\odot} (m_{J}/1 KeV)^{-2}
\end{equation}
The corresponding linear scale
$\lambda_{gal} = 2.9 \times 10^{24} h^{-2/3}$ cm.
It enters the horizon at $t = t_{H} = 2.8 \times 10^{8} h^{-4/3}$ sec
\footnote{We use for the equilibrium time
$t_{eq} = 4.49 \times 10^{10} (\Omega_0 h^2)^{-2} T_{2.73}^6$
sec. and
$1+z_{eq}=2.77 \times 10^4 (\Omega_0 h^2) T_{2.73}^{-4}$,
calculated for two light and one heavy neutrinos.}.
At this moment the amplitude of fluctuation is equal to
\beq
{(\frac{\delta \rho}{\rho})}_{\lambda_{gal}} = \delta_{H}
\eeq
which is the same for all scales entering the horizon,
for the Harrison-Zeldovich spectrum. From the moment
$t_{H}$ up to $t = t_{eq}$ the amplitude grows as
\cite{KT}
\begin{equation}
\delta_{\lambda_{gal}}(t) \equiv
{(\frac{\delta \rho}{\rho})}_{\lambda_{gal}} (t)
= \delta_{H} (1 + A \ln t / t_{H}),
\end{equation}
where $A$ can be found from the value of
$\dot \delta_{\lambda_{gal}}(t)$ at $t \leq t_{H}$.
Since for this interval
$\delta_{\lambda_{gal}}(t) = \delta_{H} (t/t_{H})$,
equating the derivatives at $t = t_{H}$ one finds
$A = 1$. From $\delta (t) \sim t^{2/3}$ for
$t \geq t_H$ it follows that
\begin{equation}
\label{xx}
{(\frac{\delta \rho}{\rho})}_{k} (t_0)
= \delta_H (1 + \ln t_{eq}/t_H) (t/t_{eq})^{2/3}
\end{equation}
where $k=2\pi/\lambda_{gal}$ is the comoving wave
number.

We now calculate the value of $\delta_{H}$ from the
quadrupole momentum of microwave radiation found
in the COBE experiment \cite{cobe} $Q = 17 \times 10^{-6}$ K
for the case of the Harrison-Zeldovich spectrum.
As shown in ref. \cite{Peebles}
${(\frac{\delta \rho}{\rho})}_\lambda$
for the comoving scale $\lambda$ is related to
the dimensionless quadrupole momentum $a_2$ as
\beq
{(\frac{\delta \rho}{\rho})}_\lambda
= \left(\frac{108}{\pi}\right)^{1/2}
a_{2} \left( \frac{c H^{-1}}{\lambda} \right)^2 I (\lambda)
\eeq
where H is the Hubble constant, $I(\lambda)=1/6$
for large scales $\lambda$, and $a_2$ is connected to Q as
\beq
a_2^2 = \frac{4\pi}{5} \frac{Q^2} {T^2}
\eeq
Using ${(\frac{\delta \rho}{\rho})}_\lambda = \delta_H$
for $\lambda = \frac{2}{3}c H^{-1}$ we obtain
\beq
\delta_{H} = 2.15 \times 10^{-5}
\eeq
Finally, for the galactic scale
$\lambda_{gal}$ we obtain from \eq{xx}
${(\frac{\delta \rho}{\rho})}_{\lambda_{gal}}(t_0) =3.6$
for h=1 and
${(\frac{\delta \rho}{\rho})}_{\lambda_{gal}}(t_0) =1.2$
for h=0.5, values which are sufficient for entering
the nonlinear stage of the perturbation's growth.

\section{A Particle Physics Model}

{}From the above discussion we see the need
to have very small couplings of the majoron
to neutrinos (so as to provide $\tau_{J} > t_0$)
as well as to charged leptons (so as to avoid too
many photons from the $J \ra \gamma \gamma$ decay).
These constraints are summarized in \eq{21} and \eq{22}.
Here we demonstrate that a realistic model
can meet such requirements in a natural way.
In our example the required smallness of the
couplings of the majoron to neutrinos and
charged leptons arises from their radiative
origin. This is achieved without the need to
introduce new mass scales, much larger than
that of the standard electroweak model.

The model contains, in addition to the Standard
Model particles, one singly charged singlet scalar $\eta^-$
\cite{zee} and one doubly charged singlet scalar $\chi^{--}$
\cite{box}. It is characterized by the following lepton Yukawa
interactions \cite{Babu88}
\beq
{\cal L} = -\frac{\sqrt2 m_i}{v}{\bar{\ell}}_i \phi e_{Ri} +
f_{ij} \ell_i^T C i \tau_2 \ell_j \eta^+ +
h_{ij} e_{Ri}^T C e_{Rj} \chi^{++} + H.c.
\label{yuk}
\eeq
($h$ and $f$ are symmetric and anti-symmetric
coupling matrices, respectively). In addition, the scalar boson
lagrangean is given by \cite{Chang,BGLAST}
\begin{eqnarray}
\label{pot1}
V_1&=&
-\frac{1}{2}\mu_\phi^2 \phi ^\dagger \phi + \frac{1}{4}\lambda_\phi
(\phi ^\dagger \phi)^2
+ \frac{1}{2}\mu_\chi^2 \chi ^\dagger \chi +
\frac{1}{4}\lambda_\chi (\chi ^\dagger \chi)^2
+\frac{1}{2}\mu_\eta^2 \eta ^\dagger \eta + \frac{1}{4}\lambda_\eta
(\eta ^\dagger \eta)^2 \nonumber\\&&
+\frac{1}{2}\epsilon_{\phi \chi } \phi ^\dagger \phi \chi ^\dagger
\chi
+\frac{1}{2}\epsilon_{\phi \eta } \phi ^\dagger \phi \eta ^\dagger \eta
+\frac{1}{2}\epsilon_{\eta \chi } \eta^\dagger \eta \chi^\dagger \chi
+ \epsilon \sigma \chi^{++} \eta^- \eta^- + H.c.
\end{eqnarray}
where $\epsilon$ and $\lambda$ are dimensionless coupling
constants and $\mu$'s are mass parameters.

To this we add
\begin{eqnarray}
\label{pot2}
V_2&=&
-\frac{1}{2}\mu_{\sigma}^2\sigma^{\dagger}\sigma
+ \frac{1}{4}\lambda_\sigma (\sigma^{\dagger}\sigma)^2
+ \epsilon_{\phi \sigma} (\phi^{\dagger}\phi)(\sigma^{\dagger}\sigma)
+ \epsilon_{\chi \sigma} (\chi^{\dagger}\chi)(\sigma^{\dagger}\sigma)
+ \epsilon_{\eta \sigma} (\eta^{\dagger}\eta )(\sigma^{\dagger}\sigma)
\end{eqnarray}
Minimizing the total potential one finds the Higgs boson
vacuum expectation value responsible for electroweak
breaking, given by
\begin{eqnarray}
\label{}
\VEV{\phi} & = & \frac{\mu_\phi }{\sqrt{\lambda _\phi }}
\end{eqnarray}
In addition, the complex singlet $\sigma$
also acquires a nonzero vacuum expectation value
$\VEV{\sigma}$ which breaks the global symmetry,
while the $\chi$ and $\eta$ mass terms are chosen positive
in such a way that the electrically charged Higgs bosons
do not acquire vacuum expectation values, as required.

The charged lepton masses are generated when the \21
symmetry is broken by $\VEV{\phi}$, through the
first term. On the other hand, neutrinos acquire
masses only radiatively, at the two loop level,
by the diagram in Fig. 2 \cite{Babu88}. This
follows directly from the spontaneous violation
of the lepton number symmetry by the nonzero
vacuum expectation value of the $\sigma$ field.

Notice that in the model summarized above the majoron
is introduced at the electroweak scale. Many variant
models of this type can be considered \cite{JoshipuraValle92}.
Such models have recently been considered in more detail
in connection with the question of electroweak baryogenesis
\cite{BGLAST} and Higgs boson physics
\cite{JoshipuraValle92}.  In all of
them, the presence of the neutral singlet $\sigma$
provides two additional electrically neutral
degrees of freedom: an extra CP-even Higgs boson
(corresponding to the $\rho$ field) and a CP-odd
scalar boson, i.e. the majoron, $J$, which
remains massless as a result of the
spontaneous nature of the lepton number
breaking. Both CP-even mass eigenstate
scalar bosons $H_1$ and $H_2$ (corresponding
to the \sm Higgs boson and the extra one coming from $\rho$)
can decay by majoron emission, either directly, such as
$H_1 \ra JJ$ and $H_2 \ra JJ$ or indirectly, via $H_2 \ra H_1 H_1$.
As long as the scale at which the lepton number breaking
is sufficiently low, as assumed here, these dark Higgs decay
modes will be substantial, suppressing the standard Higgs decays,
such as into $b\bar{b}$. As discussed above it also implies that
the $\rho$ bosons produced at the phase transition
immediately decay to the majorons that form the dark matter.

The implications of this model have recently attracted
a lot of attention \cite{JoshipuraValle92,alfonso,HJJ}.
In addition to the presence of the invisible Higgs
decay modes the production of the two
CP-even Higgs bosons through the Bjorken process
$e^+ e^- \ra Z H_i$ will be smaller than predicted
in the \sm thus weakening the LEP limit
on Higgs boson mass \cite{alfonso}.
An implication of this fact is that this model
bypasses the conflict with electroweak
baryogenesis \cite{BGLAST} which requires
$m_H \lsim 40$ GeV \cite{Dineetal92a}.

In what follows we show how this model provides
a way where all the required conditions \eq{21},
\eq{22} can be realized.

First note that the majoron coupling to neutrinos
arises from the same two-loop diagram as Fig. 2.
An estimate of this graph gives \cite{DARK92}
\begin{eqnarray}
\label{21a}
h_\nu^{ij} \approx
\frac{\epsilon \sum_{a,b} f_{ai} f_{jb} h_{ab} m_a m_b }
{256\pi^4 {M_0}^2}
\end{eqnarray}
For natural choices of parameters,
consistent with all present observations,
e.g. $f_{e\tau},  f_{\mu\tau}, h_{\tau\tau} \sim 10^{-2}$,
$\epsilon^2 \sim 10^{-5}$, and charged Higgs  boson masses
of about 200 GeV, these $h_\nu$ couplings have a typical
magnitude $h_\nu \sim 10^{-17}$ which is quite compatible
with what is required by \eq{21}.

On the other hand the corresponding
\neu masses can be written as
$$
m_\nu = h_\nu \VEV{\sigma}
$$
and lie in the $10^{-6}$ eV range or below,
for $\VEV{\sigma} \simeq 100$ GeV. Such
small \neu masses might play a role in the
understanding of the solar \neu deficit.

On the other hand, the majoron coupling to charged
leptons arises from the two-loop diagrams shown
as Fig. 3. Using the estimate from ref. \cite{Chang}
\beq
\label{22a}
h_l \sim {\frac{1}{16\pi^2}}^2
{\frac{\epsilon^2 \VEV{\phi} m_l}{M^2}}
\sum_k {(f_{lk}^2 + 2 h_{lk}^2)}
\eeq
one finds that, for reasonable choices of parameters,
$h_l$ values can lie in a range quite consistent with
what is required by \eq{22}. For example,
for the choice of parameters described
above, one gets for $h_l \sim 2 \times 10^{-13}$
for the case $l=\tau$.

Note that, although both $h_\nu$ and $h_l$ arise
at the same two-loop level, there is enough
freedom in the choice of model parameters to
obtain the required hierarchy $h_l \gg h_\nu$.

As a final comment we note that in this
model the decay $J\ra \gamma\gamma$ also
receives contributions from two-loop diagrams
mediated by electrically charged scalar bosons
(Fig. 1(b)). A rough estimate shows that it may
be of similar importance as the effective triangle
graph of Fig. 1(a).

\section{Conclusions}

In this paper we outlined the possibility of the
KeV majoron playing the role of DMP.

In order to be DMP the majoron lifetime must be longer
than the age of the universe and its effective couplings
with neutrinos and charged leptons, $h_{\nu}$ and $h_{l}$,
respectively, must be rather small. This can be naturally
realized in models where the majoron interacts with
these particles only radiatively, via the exchange
of DIPS. We have shown explicitly that in this case
these constants can be very small, as in the model
discussed in section 3.

We considered two possible scenarios for the cosmological
production of the majorons. If the DIPS are lighter than
$V_{s}$, $m_{DIP} < V_{s}$, the majorons can be for a short
time period in thermal equilibrium.  In the alternative case
the DIPS are heavier than $V_{s}$ and therefore they are not
present at the phase transition $(T \sim V_{s})$ when the
majorons are born. Since both effective couplings, $h_{\nu}$
and $h_{l}$ are very small, in this case never in the history
of the Universe were the majorons in thermal equilibrium.

As we demonstrated, this scenario is not in conflict
with nucleosynthesis data and results in a reasonable
growth factor for the density perturbations in agreement
with the limits of the COBE experiment.

A possible observational signature for our scenario is the
existence of a monochromatic line $E_\gamma = 0.5 m_J$
KeV in the extragalactic diffuse flux. Its intensity
is determined by the coupling $h_\tau$ of the majoron
to the tau lepton and equals to
\beq
I_\gamma = \frac{1}{4\pi}
\frac{\rho_{cr}}{m_J}
\frac{c t_0}{\tau_{J\gamma\gamma}}
\simeq 6 m_J^2 \left(\frac{h_\tau}{3\times 10^{-13}}\right)^2
cm^{-2} sec^{-1} sr^{-1}
\eeq
This line may be detected also from the galactic center
and nearby cluster of galaxies. We shall discuss this
problem somewhere else.
As we have already mentioned, the decay
$J\ra \gamma\gamma$ due to electrically
charged scalar boson loops (Fig. 1(b))
can compete with the process discussed above.

A particle physics signature for this model
(section 3) consists in the possibility of
observable rates for flavour violating decays
such as $\mu \ra e \gamma$ and $\mu \ra 3e$.
The first decay process arises at one-loop,
due to singly-charged scalar boson
exchange, while the second is present already
at the tree level, due to $\chi^{++}$ exchange.
These processes could be tested at laboratory
experiments such as MEGA or at PSI.

\vfill
This work was supported by DGICYT under grant
PB92-0084 and by Generalitat Valenciana.
We thank A. Dolgov, I. Khriplovich, J. Peltoniemi
and A. Smirnov for valuable discussions.

\newpage

\newpage
\section*{Figure Captions}
\noindent
{\bf Fig. 1. }\\
Diagrams leading to the decay $J \ra \gamma \gamma$:
(a) Effective triangle loop contributions;
(b) Scalar boson mediated contributions.\\
{\bf Fig. 2.}\\
Diagram generating the small coupling $h_\nu$ of the
majoron to neutrinos. It also generates very small
Majorana neutrino masses\\
{\bf Fig. 3.}\\
Diagram generating the coupling $h_l$ of the majoron
to charged leptons.\\

\end{document}